\documentclass[11pt]{article}
\usepackage{amssymbols}
\usepackage{fullpage}

\makeatletter

\newtheorem{theorem}{Theorem}[section]

\newtheorem{observation}[theorem]{Observation}

\newcommand{\tuple}[1]{\langle #1 \rangle}
\newcommand{\NP}{\mathbf{NP}}

\newcommand{\SP}{\mathbf{\# P}}
\renewcommand{\P}{\mathbf{P}}

\newcommand{\PP}{\mathbf{PP}}
\newcommand{\PSPACE}{\mathbf{PSPACE}}
\newcommand{\BPP}{\mathbf{BPP}}
\newcommand{\BQP}{\mathbf{BQP}}
\newcommand{\EQP}{\mathbf{EQP}}
\newcommand{\QP}{\mathbf{QP}}
\newcommand{\AWPP}{\mathbf{AWPP}}

\begin{document}

\title{One Complexity Theorist's View of Quantum Computing%
  \thanks{Based on a talk presented at the Second Workshop on
    Algorithms in Quantum Information Processing at DePaul University,
    Chicago, January 21, 1999. Realaudio of the talk is available at http://www.cs.depaul.edu/aqip99.}}

\author{Lance Fortnow\thanks{URL:
http://www.neci.nj.nec.com/homepages/fortnow. Email:
fortnow@research.nj.nec.com.}\\NEC Research Institute\\4 Independence
Way\\Princeton, NJ 08540}

\maketitle

\begin{abstract}
The complexity of quantum computation remains poorly understood. While
physicists attempt to find ways to create quantum computers, we still
do not have much evidence one way or the other as to how useful these
machines will be. The tools of computational complexity theory should
come to bear on these important questions.

Quantum computing often scares away many potential researchers from
computer science because of the apparent background need in quantum
mechanics and the alien looking notation used in papers on the topic.

This paper will give an overview of quantum computation from the point
of view of a complexity theorist. We will see that one can think of
$\BQP$ as yet another complexity class and study its power without
focusing on the physical aspects behind it.
\end{abstract}

\section{Introduction}

When one starts looking at quantum computation an obvious question
comes to mind: {\em Can quantum computers be built?} By this I mean
can one create a large reliable quantum computer---perhaps to factor
numbers using Shor's algorithm~\cite{Shor} faster than any classical
computer can factor.  There has been some success creating in creating
quantum machines with a tiny number of bits, we have many physical and
engineering issues to overcome before a large scale quantum machine
can be realized.

As a computational complexity theorist, I would like to consider a
different and perhaps more important question: {\em Are quantum
  computers useful?} In other words, even if we can build reasonable
quantum machines, will they actually help us solve important
computational problems not attainable by traditional machines. Realize
that the jury is still out on this issue. Grover's
algorithm~\cite{Grover} gives only a quadratic speed-up, which means
that a quantum machine will have to approach speeds of a traditional
computer (on the order of a trillion operations per second) before
Grover overtakes brute-force searching.

Shor's quantum algorithms for factoring and discrete
logarithm~\cite{Shor} give us better examples. Here we achieve an
exponential speed-up from the best known probabilistic algorithms.
Factoring alone does not make quantum computing useful.  Consider the
following paragraph from Shor's famous paper~\cite[p.\ 1506]{Shor}:
\begin{quote}
  Discrete logarithms and factoring are not in themselves widely
  useful problems.  They have become useful only because they have
  been found to be crucial for public-key cryptography, and this
  application is in turn possible only because they have been presumed
  to be difficult. This is also true of the generalizations of Boneh
  and Lipton~\cite{Boneh-Lipton} of these algorithms. If the only uses
  of quantum computation remain discrete logarithms and factoring, it
  will likely become a special-purpose technique whose only raison
  d'etre is to thwart public key cryptosystems.
\end{quote}

How useful can quantum computing be? What is the computational
complexity of $\BQP$, the class of efficiently quantum computable
problems. Does $\BQP$ contain $\NP$-complete problems like graph
3-colorability? What is the relationship of $\BQP$ and other more
traditional complexity classes like the polynomial-time hierarchy?

These questions are well suited for computational complexity
theorists. Many researchers though shy away from quantum complexity as
it appears that a deep knowledge of physics is necessary to understand
$\BQP$. Also most papers use a strange form of notation that make
their results appear more difficult than they really are.

This paper argues that one can think about $\BQP$ as just a regular
complexity class, not conceptually much different than its
probabilistic cousin $\BPP$. We develop a way of looking at
computation as matrix multiplications and exhibit some immediate
effects of this formulation. We show that $\BQP$ is a rather robust
complexity class that has much in common with other
classes. We argue that $\BQP$ and $\BPP$ exhibit much of the same
behavior particularly in their ability to do searching and hiding of
information.

While this paper does not assume any knowledge of physics, we do
recommend a familiarity with basic notions of complexity theory
as described in Balc{\'a}zar, D\'{\i}az and Gabarr{\'o}~\cite{BDG}.

\section{Computation as Matrix Multiplication}
\label{matrixsec}

Consider a multitape deterministic Turing machine $M$ that uses $t(n)$
time and $s(n)$ space. Let us also fix some input $x$ in $\{0,1\}^n$
and let $t=t(n)$ and $s=s(n)$. We will assume $t(n)\geq n$ and
$s(n)\geq \log n$ are fully time and space constructible
respectively. We will always have $t(n)\leq s(n)\leq 2^{O(t(n))}$.

Recall that a configuration $c$ of machine $M$ on input $x$ consists
of the contents of the work tapes, tape pointers and current state. We
will let $C$ be the number of all configurations of $M$ on input
$x$. Note that a read-only input tape is not considered as part of the
configuration and thus we have $C=2^{O(s)}$.

For most of the paper, we will refer to the polynomial-time case,
where $t(n)=n^{O(1)}$ and $s(n)=2^{t(n)}=2^{n^{O(1)}}$.

Fix an input $x$. Let $c_I$ be the initial configuration of
$M(x)$. We can assume without loss of generality that $M$ has exactly
one accepting configuration $c_A$ and that once $M$ enters $c_A$ it
remains in $c_A$.

Define a $C\times C$ transition matrix $T$ from the description 
of the transition function of $M$. Let $T(c_a,c_b)=1$ if one can get
from $c_a$ to $c_b$ in one step and $T(c_a,c_b)=0$ otherwise.

\begin{observation}
\label{obsdet}
For any two configurations $c_a$ and $c_b$, $T^r(c_a,c_b)=1$ if and
only if $M$ on input $x$ in configuration $c_a$ when run for $r$ steps 
will be in configuration $c_b$.
\end{observation}
In particular we have that $M(x)$ accepts if and only if
$T^t(c_I,c_A)=1$.

Now what happens if we allow $M$ to be a nondeterministic machine. Let
us formally describe how the transition function $\delta$ works for a
$k$-tape machine (along the lines of Hopcroft and Ullman~\cite{HoUl}).
\[\delta:Q\times\Sigma\times\Gamma^k\times
Q\times\Gamma^k\times\{L,R\}^{k+1}\rightarrow\{0,1\}\] Here $Q$ is the
set of states, $\Sigma$ is the input alphabet and $\Gamma$ the tape
alphabet. The transition function
$\delta(q,a,b_1,\ldots,b_k,p,c_1,\ldots,c_k,d_0,d_1,\ldots,d_k)$ takes
the value one if the transition from state $q$ looking at input bit
$a$ and work tape bits $b_1$, $b_2$, $\ldots$, $b_k$ moving to state $p$,
writing $c_1$, $\ldots$, $c_k$ on the work tapes, moving the input head
in the direction $d_0$ and the tape heads in directions $d_1$,
$\ldots$, $d_k$ respectively is legal.

This $\delta$ function yields again a transition matrix $T$ in the
obvious way: If it is possible to reach $c_b$ from $c_a$ in one step
then $T(c_a,c_b)=1$. If it is not possible to go from $c_a$ to $c_b$
in one transition then $T(c_a,c_b)=0$.

A few notes about the matrix $T$: Most entries are zero. For
polynomial-time computation, given $c_a$, $c_b$ and $x$ one can in
polynomial-time compute $T(c_a,c_b)$, even though the whole transition 
matrix is too big to write down in polynomial-time.

Observation~\ref{obsdet} has a simple generalization:
\begin{observation}
\label{obsnondet}
For any two configurations $c_a$ and $c_b$, $T^r(c_a,c_b)$ is the
number of computation paths from $c_a$ to $c_b$ of length $r$.
\end{observation}
We now have that $M$ on input $x$ accepts if and only if $T^t(c_I,c_A)>0$.

We can define $\#M(x)=T^t(c_I,c_A)$ as the number of accepting paths
of machine $M$ on input $x$.

For polynomial-time machines this yields the $\SP$ functions first
defined by Valiant~\cite{Va}.

Now interesting events happen when we allow our $\delta$ function to
take on nonbinary values. First let us consider $\delta$ taking on
nonnegative integers. We get a further generalization of
Observations~\ref{obsdet} and~\ref{obsnondet}:
\begin{observation}
\label{obsgen}
The value of $T^r(c_a,c_b)$ is the sum over all computation paths from 
$c_a$ to $c_b$ of the product of the values of $\delta$ over each
transition on the path.
\end{observation}

What kind of functions do we get from $T^t(c_I,c_A)$ for
polynomial-time machines? We still get exactly the $\SP$
functions. Suppose that we have a $T(c_a,c_b)=k$. We simply create a
new nondeterministic Turing machine that will have $k$ computation
paths, of length about $\log k$, from $c_a$ to $c_b$. This will only
increase the running time by a constant factor.

Now suppose we allow our $\delta$ function to take on nonnegative
rational values. Remember that the $\delta$ function is a finite
object independent of the input. Let $v$ be the least common multiple
of the denominators of all the possible values of the $\delta$
function. Define $\delta_*=v\delta$.  Let $T$ and $T_*$ be the
corresponding matrices for $\delta$ and $\delta_*$ respective. Note $T_*=vT$.

The values $T_*^t$ still capture the $\SP$ functions. We also have
$T^t=T_*^t/v^t$. Since $v^t$ is easily computable this does not give
us much more power than before.

We can use a restricted version of this model to capture probabilistic
machines. Probabilistic machines have 
\[\delta:Q\times\Sigma\times\Gamma^k\times
Q\times\Gamma^k\times\{L,R\}^{k+1}\rightarrow[0,1]\]
with the added restriction
that for any fixed values of $q$, $a$ and $b_1,\ldots,b_k$,
\[\sum_{p,c_1,\ldots,c_k,d_0,d_1,\ldots,d_k}
\delta(q,a,b_1,\ldots,b_k,p,c_1,\ldots,c_k,d_0,d_1,\ldots,d_k)=1.\]

What does this do to the matrices $T$? Every row of $T$ sums up to
exactly one and its entries are all range in between zero and one,
i.e., the matrices are {\em stochastic}.  Note also that stochastic
matrices preserve the $L_1$ norm, i.e., for every vector $u$,
$L_1(u\cdot T)=L_1(u)$. These observations will be useful to us as we
try to understand quantum computation.

Now $T^t(c_I,c_A)$ computes exactly the probability of acceptance of
our probabilistic machine. Keep in mind that $T$ still is mostly zero
and although the matrix $T$ is exponential size, the entries are
easily computable given the indices.

We can define the class $\BPP$: A language $L$ is in $\BPP$ if there
is a probabilistic matrix $T$ as described above and a polynomial $t$
such that
\begin{itemize}
\item For $x$ in $L$, we have $T^t(c_I,c_A)\geq 2/3$.
\item For $x$ not in $L$, we have $T^t(c_I,c_A)\leq 1/3$.
\end{itemize}
Using standard repetition tricks to reduce the error we can replace
$2/3$ and $1/3$ above with $1-2^{-p(|x|)}$ and $2^{-p(|x|)}$ for any
polynomial $p$.

Let us make some variations to the probabilistic model. First let us
allow $\delta$ to take on arbitrary rational values including
negative values. Since the value $T^t$ may be negative we will use
$(T^t(c_I,c_A))^2$ as our ``probability of acceptance''. Finally
instead of preserving the $L_1$ norm, we preserve the $L_2$ norm
instead. That is for every vector $u$, we have
\[L_2(T(u))=L_2(u)=\sqrt{\sum_a u_a^2}.\]
Unlike the $L_1$ norm, preserving the $L_2$ norm is a global
condition, we cannot just require that the squares of the values of
the $\delta$ function add up to one. Matrices that preserve the $L_2$
norm are called {\em unitary}.

This model looks strange but still rather simple. Yet this model
captures exactly the power of quantum computing!

We can define the class $\BQP$: A language $L$ is in $\BQP$ if there
is a quantum matrix $T$ as described above and a polynomial $t$
such that
\begin{itemize}
\item For $x$ in $L$, we have $(T^t(c_I,c_A))^2\geq 2/3$.
\item For $x$ not in $L$, we have $(T^t(c_I,c_A))^2\leq 1/3$.
\end{itemize}

\section{Questions}
\label{questionsec}

For those who have seen quantum computing talks in the past, many
questions may come up.

%%\begin{itemize}
%%\item Where's the physics?
%%\item Where's the $<$bra$|$ and $|$ket$>$ notation?
%%\item Don't we need complex numbers to do quantum computation?
%%\item Don't we have to require the computation to be reversible?
%%\item What if there are many accepting configurations?
%%\item Don't we have to take measurements?
%%\end{itemize}

\subsection{Where's the physics?}

The presentation of quantum computing above is based mostly on the
computation model developed by Bernstein and Vazirani~\cite{BV}. We
have presented the model here as another complexity model. It does
however capture all of the physical power and rules of quantum
computation.

Can one study quantum computation without a deep background in quantum
mechanics? I say yes. I drive a car without much of a clue as to what
a carburetor does. I can program a classical computer and do research
on the computational complexity of these computers even though I do
not have a real understanding of how a transistor works. I often do
research on theoretical computation models such as nondeterminism that
have no physical counterpart at all. Given a good model of a quantum
computer, computer scientists can study its computational abilities
without much knowledge of the physical properties of that model.

\subsection{Where's the $\langle$bra$|$ and $|$ket$\rangle$ notation?}

When researchers in two disciplines try to work they usually find that
language forms a major wall between them. Not the base spoken
language, but the notation and definitions each one assumes that every
first-year graduate student in the field knows and follows. In quantum
computing, generally physics notation has prevailed, leading to one of
the major barriers to entering the area.

Physicists generally use the Dirac bra-ket notation to represent base quantum
states. When computer scientists see this notation, it looks quite
foreign and many assume that the complexity of the notation reflects
some deep principles in quantum mechanics. Nothing could be further
from the truth as the bra-ket notation represents nothing more than
vectors.

The notation $|010\rangle$ is called a ``ket'' and represents the
string $010$. In general $|x\rangle$ with $x\in\{0,1\}^k$ represents
the string $x$.  Concatenation is concatenation:
$|x\rangle|y\rangle=|xy\rangle$.  The strings of length $k$ form the
basis of a vector space of dimension $2^k$. These basis vectors
roughly correspond to the configurations described in
Section~\ref{matrixsec}.

Now, there is also a ``bra'' and various rules that apply
when putting bras and kets together but these rarely come into play in 
quantum computing. For the interested reader there are many sources
one can read on the subject such as Preskill's lecture 
notes~\cite{Preskill-notes}.

Why does the simple looking notation cause such confusion for computer
scientists? This notation does not conform to computer science
conventions, particular to that of symmetry. When we consider vectors
or other groupings of objects, computer scientists always use
symmetric brackets such at $\tuple{8,3,4}$, $[0\ldots 1]$, $(a+b)^*$
and $|x|$.  When typical computer scientists look at a character like
``$\rangle$'' or ``$|$'' that does not have a counterpart, they
consider these as relational operators, making equations like
\begin{eqnarray}
\label{noteex}
|\psi_1(y)\rangle = \frac{1}{\sqrt{2^n}}\sum_x(-1)^{\delta_{x,y}}
|x\rangle =|\psi_0\rangle-\frac{2}{\sqrt{2^n}}|y\rangle & \ &
\mbox{\cite[p.\ 1514]{BBBV}}
\end{eqnarray}
impossible even to parse.

As this paper suggests, one does not need to know Dirac notation to
understand the basics of quantum computation. Unfortunately one is
forced to learn it to follow the communication in the field.

If theoretical computer scientists collaborate then they must learn
\LaTeX. It does not matter if they like some other mathematical
document software, they have to use \LaTeX\ if they wish to write
papers with other computer scientists.

Likewise if you want to truly study quantum computation you will need
to learn the Dirac notation. One can easily develop more natural
notation for computer scientists but we have already lost the
battle. Eventually the notation becomes easier to parse and follow and
formulas like Equation~(\ref{noteex}) seem almost normal. I find it a
shame though that computer scientists find it difficult to go to talks
on quantum computing not so much because of the complexity of the
subject matter but because of the abnormal notation.

\subsection{Don't we need arbitrary real and complex numbers to do
quantum computation?}

In short no. The physics definition allows arbitrary real and complex
entries as long as the matrices are unitary. This by itself limits the
values to have absolute value at most one. But one can do much better.
%The key thing to remember is that we only need to approximate the
%accepting probability, as we shall see later, to achieve the same
%class of $\BQP$.

If we allow all possible reals, $\BQP$ can accept arbitrarily
complicated languages~\cite{ADH}. However this result feels like
cheating---basically you encode hard languages directly into the
entries of the matrix. Thus one requires knowing the language ahead
of time to create the machine. A similar trick can also be played with
probabilistic machines using noncomputable probabilities.

For fairness we should only allow efficiently computable matrix entries,
where we can compute the $i$th bit in time polynomial in $i$.
Independently Adleman, DeMarrais and Huang~\cite{ADH} and Solovay and
Yao~\cite{SoYa} show that we can simulate a $\BQP$ machine using
efficiently computable entries from the set
$\{-1,-\frac{4}{5},-\frac{3}{5},0,\frac{3}{5},\frac{4}{5},1\}$.  Or
you can get away with fewer numbers if you don't mind an irrational:
$\{-1,-\sqrt{2},0,\sqrt{2},1\}$.

\subsection{Don't we have to require the computation to be reversible?}

The matrix $T$ as well as any power of $T$ is unitary, i.e. it
preserves the $L_2$ norm. For real valued matrices, $T$ is unitary if
and only if the transpose of $T$ is the inverse of $T$. The reader
should try the proof himself or it can be found in Bernstein and
Vazirani~\cite[p.\ 1463]{BV}.

In particular this means that $T$ is invertible. Even more so the
inverse of $T$ is itself unitary. If you have a vector $v$ over the
configurations that gives the value for each configuration, let
$w=T(v)$. We then have $T^{-1}(w)=v$. Given the current entire
state of the system we can reverse it to get the previous state.

Don't think of reversibility as a requirement of quantum computing
rather as a consequence. Nevertheless keep in mind that in creating a
quantum algorithm at a minimum you will need to create a reversible
procedure.

\subsection{What if there are many accepting configurations?}

For traditional models of computation we can usually assume only one
accepting configuration. We do this by a cleanup operation---after the
machine decides to accept we erase the work tapes, move the head to
the left and enter a specified final accepting state.

This procedure will not work for quantum computation since it is not
reversible. Bennett, Bernstein, Brassard and Vazirani~\cite{BBBV} show
an interesting way to get around this problem: Compute whether the
quantum machine accepts. Copy the answer to an extra quantum bit. Then
reverse the whole process. The unique accepting state is now the
initial state with this extra bit turned on.

\subsection{Don't we have to allow measurements?}
\label{meassec}

Our formulation of $\BQP$ does not appear to contain any
measurements. In fact we are simulating a measurement at the end. A
$\BQP$ machine will output ``accept'' with probability equal to the
sum of the squares of the values of the accepting
configurations. We have only one accepting configuration so we need
just consider $(T^t(c_I,c_A))^2$.

One could consider a process that allows measurements during the
computation. Bernstein and Vazirani~\cite{BV} show that we can push
all of the measurements to the end. Basically measurements are linear
projections and we can simulate a measurement by doing a rotation and
taking one big projection at the end.

\section{Full Robustness}
Section~\ref{questionsec} points to the fact that $\BQP$ is quite a
robust complexity class. Bernstein and Vazirani~\cite{BV} show two
more important facts: That $\BQP$ can simulate any deterministic or
probabilistic polynomial-time algorithm and $\BQP^\BQP=\BQP$. In other
words we can do quantum subroutines and build that directly into the
quantum computation. The proofs of these facts are technically quite
complicated but here a few points to think about.

Note that a deterministic computation might not be reversible but
Bennett~\cite{Bennett-rev} showed how to simulate any deterministic
computation by a reversible computation with only squaring the
computation time.

Flipping a truly random coin seems inherently nonreversible: Once
flipped how does one recover the previous state? Quantum computing
does have one nonreversible operation known as measurements. One can
simulate truly random coin tosses using measurements, or as described
in Section~\ref{meassec} simulating a measurement and doing it at the end.

From these results we can use that standard error reduction techniques
to reduce the error in a quantum algorithm to $2^{-p(n)}$ where $p$ is
a polynomial and $n$ is in the length of the input.

Other complexity classes such as the zero-error version of $\BQP$
(denoted $\EQP$ or $\QP$) may not have some of the nice robust
properties of Section~\ref{questionsec} since approximating the
matrix entries would no longer keep us in the class.

Space-bounded classes also lack some of the robustness of $\BQP$. For
example if we do not allow a measurement until the end even simulating
coin flips becomes difficult: we cannot reuse the coins and keep
reversibility. See Watrous~\cite{Watrous-random} for a discussion.

\section{Using This Formulation}

Once you have the formulation of quantum machines described in
Section~\ref{matrixsec}, one immediately gets interesting results on
quantum complexity.

\subsection{$\AWPP$}
\label{awppsec}

Suppose we remove the unitary restriction in the definition of $\BQP$
in Section~\ref{matrixsec}. This yields the class $\AWPP$.
Li~\cite{Li-PhD} and Fenner, Fortnow, Kurtz and Li~\cite{FFKL} defined
and studied the class $\AWPP$ (stands for Almost-Wide Probabilistic
Polynomial-time) extensively. They showed many interesting results for
this class.
\begin{enumerate}
\item $\AWPP\subseteq\PP\subseteq\PSPACE$, where $\PP$ is the set of
languages accepted by a probabilistic polynomial-time Turing machine
where we only require the error probability to be smaller than
one-half.
\item The class $\AWPP$ is low for $\PP$, i.e., for any language $L$
in $\AWPP$, $\PP^L=\PP$.
\item If $\P=\PSPACE$ then relative to any generic oracle $G$,
$\P^G=\AWPP^G$. This implies that without any assumption there is a
relativized world $A$ such that $\P^A=\AWPP^A$ and the polynomial-time
hierarchy is infinite.
\item There exists a relativized world where $\AWPP$ does not have
complete sets, in fact where $\AWPP$ has no sets hard for all of $\BPP$.
\end{enumerate}
Relativization results show us important limitations on how certain
techniques will solve problems in complexity theory. For a background
on relativization see the survey paper by
Fortnow~\cite{F-relative}.

Fortnow and Rogers~\cite{FR-quantum} observed that
$\BQP\subseteq\AWPP$ basically falls out of the characterization given
in Section~\ref{matrixsec}. From this we have several corollaries.
\begin{enumerate}
\item $\BQP\subseteq\PP\subseteq\PSPACE$ \cite{ADH}
\item The class $\BQP$ is low for $\PP$.
\item There exists a relativized world where $\P=\BQP$ and the
polynomial-time hierarchy is infinite.
\item There exists a relativized world where $\BQP$ has no complete
sets, not even any sets hard for $\BPP$.
\end{enumerate}

\subsection{Decision Trees}
The oracle results mentioned in Section~\ref{awppsec} follow from
results on decision tree complexity that have interest in their own
right. Consider the situation where we wish to compute a function from
$\{0,1\}^N$ to $\{0,1\}$ where access to the input is limited as
oracle questions. Here the complexity measure is the number of queries
needed to compute the function, not the running time.

Beals, Buhrman, Cleve, Mosca and de Wolf~\cite{BBCMW} show how to vary
the matrix model of Section~\ref{matrixsec} to create special linear
transformations that describe a reversible oracle query. One can
interleaves matrices describing these transformations with the unitary
matrices given by the computation of the Turing machine.

Grover~\cite{Grover} shows how to get a nontrivial advantage with
quantum computers: He shows that computing the OR function needs only
$O(\sqrt{N})$ queries although classically $\Omega(N)$ input bits
are needed in the worst case.

Bernstein and Vazirani~\cite{BV} give a superpolynomial gap and
Simon~\cite{Simon-quantum} gives an exponential gap. Both of
these gaps require that there are particular subsets of the inputs to
which $f$ is restricted.

Beals, Buhrman, Cleve, Mosca and de Wolf~\cite{BBCMW} notice that the
characterization given in Section~\ref{matrixsec} shows that any
quantum decision tree algorithm making $t$ queries is a polynomial in
the queries of degree at most $2t$.

Nisan and Szegedy~\cite{NS} show that if a polynomial of degree $d$
on $\{0,1\}^N$ approximates the characteristic function of some
language $L\subseteq\{0,1\}^N$ then $L$ has deterministic decision
tree complexity polynomial in $d$.

Combining these ideas, if there is a quantum algorithm computing a
function $f$ defined on all of $\{0,1\}^N$ and using $t$ queries then
there exists a deterministic algorithm computing $f$ using polynomial
in $t$ queries.

\section{Probabilistic versus Quantum}
% Entanglement
% Parallism
% Searching versus Hiding
% Negative numbers
% Unitary

Often one hears of the various ways that quantum machines have
advantages over classical computation, for example, the ability to go
into many parallel states and that states can be entangled with each
other. In fact often these properties occur in probabilistic
computation as well. The strength of quantum computing lies in the
ability to have bad computation paths eliminate each other thus
causing some good paths to occur with larger probability. The ability
for quantum machines to take advantage of this interference is
tempered by their restriction to unitary operations.

Suppose you videotape a football game from the television. Let us say
the teams are evenly matched so that either team has an equal
probability of winning. Now make a copy of the tape without watching
it. Now these videotapes are in some sense entangled. We don't know who
won the game until we watch it but both outcomes will be the same. I
can take a videotape to Mars and watch it there and at the end I will
instantaneously know the outcome of the game on the other tape.

Entanglement works quite differently for quantum computing. In
particular the videotape model fails to capture quantum
entanglement---no ``hidden variable'' theory can capture the effects
of quantum entanglement (see a discussion in
Preskill~\cite{Preskill-notes}). However we should not consider
entanglement to be of much help in the power of $\BQP$.

When it comes to parallelism there is really no difference between
probabilistic and quantum computation. Consider the value
$T^t(c_I,c_A)$ expanded as follows:
\[T^t(c_I,c_A)=\sum_{c_1,\ldots,c_{t-1}}T(c_I,c_1)T(c_1,c_2)
\cdots T(c_{t-2},c_{t-1})T(c_{t-1},c_A)\]
One can think of the vector $\tuple{c_1,\ldots,c_{t-1}}$ as describing
a computation path, the value \[T(c_I,c_1)T(c_1,c_2)
\cdots T(c_{t-2},c_{t-1})T(c_{t-1},c_A)\] as the value along that path
and the final value as the sum of the values over all paths.

Did this view of parallelism depend on whether we considered quantum
or probabilistic computation? Not a bit.

Where does the power of quantum come from? From interference. For
probabilistic computation the value
\[T(c_I,c_1)T(c_1,c_2)\cdots T(c_{t-2},c_{t-1})T(c_{t-1},c_A)\]
is always nonnegative. Once a computation path has a positive value it
is there to stay. For quantum computation the value could be negative
on some path causing cancellation or interference. This allows other
paths to occur with a higher probability. The restriction of quantum
computers to unitary transformations limits the applications of
interference but still we can achieve some useful power from it.

\subsection{Searching versus Hiding}

Randomness plays two important but quite different roles in the theory
of computation. Typically we use randomness for searching. In this
scenario, we have a large search space full of mostly good
solutions. Deterministically we can only look in a relatively small
number of places and may miss the vast majority of good solutions. By
using randomness we can, with high probability, find the good
solutions. Typical examples include primality testing~\cite{SoSt} and
searching undirected graphs in a small amount of space~\cite{AKLLR}.

However these randomness techniques appear to apply to only a small
number of problems. Recent results in derandomization indicate we
probably can do as well with deterministically chosen pseudorandom
sequences.

We also use randomness to hide information. A computer, no matter how
powerful, cannot predict the value of a truly random bit it cannot
see. Hiding forms the basis of virtually all cryptographic
protocols. Hiding also plays an important role in complexity
theory. The surprising strength of interactive proof
systems~\cite{LFKN,Sh,BFL,ALMSS} comes from the inability of the
prover to predict the verifier's future coin tosses.

We have the same dichotomy between searching and hiding in quantum
bits. For quantum searching we do not necessarily require that most
paths have high probability. The use of quantum interference allows us
to find some things with certain kinds of structure. Shor's algorithms
for factoring and discrete logarithm~\cite{Shor} form the obvious
example here.

While we do not expect that there exists any notion of quantum
pseudorandom generators, at the moment we still only have a limited
collection of problems where quantum search greatly helps over
classical methods.

Quantum hiding, like its probabilistic counterpart, has a
powerful effect on computation theory. Not only can no computer
predict the value of a quantum bit not yet measured but the quantum
bit cannot be copied and any attempt at early measurement will change
the state of the quantum bit.

Quantum bits do have one disadvantage over probabilistic bits. When we
flip a probabilistic coin the original state has been irrevocably
destroyed. The reversibility of quantum computation prevents
destroying quantum state. However, we can get around this problem by
using classical probabilistic bits to determine how to prepare the
quantum states. We can always get classical probabilistic bits by
measuring an appropriate quantum state.

Quantum hiding in this manner gives powerful tools for quantum
cryptography~\cite{BBE} and quantum interactive proof
systems~\cite{watrous-ip} that go beyond what we believe possible with
classical randomness.

\section{Future Directions}
% NP in BQP, BQP and PH, BQP and random oracles.
% Further Reading

Quantum computation is ripe for complexity theorists. There are still
many fundamental questions remaining to be solved.

Is $\NP$ in $\BQP$? While we probably cannot answer that problem
directly we could show some unlikely consequences like the
polynomial-time hierarchy collapses.

Does $\BQP$ sit in the polynomial-time hierarchy? The only evidence
against this comes from Bernstein-Vazirani~\cite{BV}. In their paper
they create a relativized world $A$ and a language $L$ that sits in
$\EQP^A\subseteq\BQP^A$ but not $\BPP^A$. We cannot prove that $L$
sits in the polynomial-hierarchy relative to $A$. However the language
$L$ does sit in deterministic quasipolynomial ($2^{\log^{O(1)}n}$)
time and by an alternating Turing machine that uses only
polylogarithmic alternations where these machines have access to
$A$. So while we may find it difficult to show $\BQP$ is in the
polynomial-time hierarchy, we could show that every language in $\BQP$
is accepted by some alternating Turing machine using polylogarithmic
alternations and/or polynomial-time.

For such results we will need more than the fact than $\BQP$ sits
in $\AWPP$ as described in Section~\ref{awppsec}. Relative to random
oracles $\NP$ is in $\AWPP$ and there exist relative worlds where an
infinite polynomial-time hierarchy sits strictly inside of
$\AWPP$. Somehow we will need to make better use of the unitary nature
of the quantum transformations.

Shor's algorithm shows that quantum machines can defeat many commonly
used cryptographic protocols. Can we create one-way functions that no
quantum machine can invert better than random guessing? Perhaps we
will need such functions computed by quantum machines as well.

Maybe quantum relates to other classes out there? Is $\BQP$ equivalent
to the problems having statistical zero-knowledge proof systems? This
seems far fetched but we have no evidence against this.

For further reading I recommend that the reader get a hold of SIAM
Journal on Computing Volume 26 Number 5 (1997). Half of this issue is
devoted to quantum computation and as one can see from the references
it has many of the important papers in the area.

Preskill's course notes~\cite{Preskill-notes} give a detailed
description of quantum computation from the physicists' point of
view. Books on the quantum computation are also starting to appear. An
early book by Gruska~\cite{Gruska} gives a broad range of work on
quantum computation.

\section*{Acknowledgments}

I wish to thank the organizers of the AQIP '99 conference for inviting
me to give the presentation and inviting me to write this paper based
on that presentation for the special issue of the conference.

I have had many exciting discussions on these topics with several top
researchers in the field including Harry Buhrman, Peter Shor, Charlie
Bennett, Gilles Brassard and Dorit Aharanov. The opinions expressed in this
paper remain solely my own.

Dieter van Melkebeek gave many useful comments on an earlier draft of
this paper.

\bibliographystyle{alpha} \bibliography{b}

\end{document}